\documentclass[letter,twocolumn]{jpsj2} 

\usepackage{times}
\usepackage{graphicx}

\title{Kondo Effect in an Electron System
with Dynamical Jahn-Teller Impurity}

\author{Takashi {\sc Hotta}}

\inst{Advanced Science Research Center,
Japan Atomic Energy Agency, Tokai, Ibaraki 319-1195}

\recdate{\today}

\abst{
We investigate how Kondo phenomenon occurs in the Anderson model
dynamically coupled with local Jahn-Teller phonons.
It is found that the total angular moment
composed of electron pseudo-spin and phonon angular moments
is screened by conduction electrons.
Namely, phonon degrees of freedom essentially contribute
to the formation of singlet ground state.
A characteristic temperature of the Kondo effect due to dynamical
Jahn-Teller phonons is explained by an effective $s$-$d$ Hamiltonian
with anisotropic exchange interaction obtained
from the Jahn-Teller-Anderson model in a non-adiabatic region.
}

\kword{dynamical Jahn-Teller effect, Kondo effect, Anderson model,
$s$-$d$ model}

\begin{document}
\maketitle


Recent discovery of exotic heavy-fermion phenomenon
in filled skutterudite compound SmOs$_4$Sb$_{12}$
\cite{Sanada,Yuhasz}
has triggered rapid increase of renewed attention to
quasi-Kondo phenomena with phononic origin.
In particular, it has been clarified that
the specific heat coefficient is almost independent of
an applied magnetic field.\cite{Sanada}
Namely, the heavy-fermion behavior in SmOs$_4$Sb$_{12}$
is magnetically robust.
It has been pointed out that this peculiar phenomenon
can be understood by non-magnetic Kondo effect
originating from phonons.\cite{Miyake}

In general, the Kondo-like phenomenon occurs
in a conduction electron system hybridized with
a localized entity with internal degrees of freedom.
When electrons are coupled with Einstein phonons,
a double-well potential is formed in an adiabatic approximation,
leading naturally to a two-level system,
in which Kondo has first considered a possibility of
non-magnetic Kondo behavior.\cite{Kondo}
The two-level Kondo system exhibits
the same behavior as the magnetic Kondo effect.\cite{Vladar}
Further intensive efforts to include the low-lying levels of local phonon
have been made to clarify the quasi-Kondo behavior
in electron-phonon systems.\cite{Miyake,Mitsumoto}

In the filled skutterudite structure, rare-earth ion is
surrounded by the cage composed of twelve pnictogens.
Then, the rare-earth ion easily moves around potential minima
in off-center positions inside the pnictogen cage.
This is called the rattling, which is considered to be
a key ingredient with significant influence on
electronic properties of filled skutterudites.
As a natural extension of the two-level Kondo problem,
Hattori {\it et al.} have analyzed the four- and six-level Kondo systems
to consider the effect of rattling in filled skutterudites.
\cite{Hattori}
Their results seem to be consistent with the magnetically robust
heavy-fermion behavior observed in SmOs$_4$Sb$_{12}$.

Concerning the problem of rattling, the measurement of
elastic constant of PrOs$_4$Sb$_{12}$ has suggested that
the off-center motion of Pr atom inside the Sb cage
has degenerate $E_{\rm g}$ symmetry.\cite{Goto}
When we attempt to include the effect of off-center motion
into a model Hamiltonian,
a simple way is to regard the rattling
as relative vibration of surrounding atoms,
leading to an electron-phonon interaction term.
In a phenomenological level, the present author has considered
a linear coupling between degenerate $f$-electron orbitals
with $e_{\rm u}$ symmetry and dynamical Jahn-Teller phonons
with $E_{\rm g}$ symmetry
in a multiorbital Anderson model constructed based on
a $j$-$j$ coupling scheme.\cite{Hotta1,Hotta2a,Hotta2b,Hotta2c}
Numerical analysis of this model has revealed
that quasi-Kondo behavior occurs
due to the release of an entropy $\log 2$ of the vibronic ground state,
originating from the clockwise and anti-clockwise rotational modes
of dynamical Jahn-Teller phonons.\cite{Hotta1}
This scenario is believed to have a potential to explain
the Kondo effect with
non-magnetic origin in filled skutterudites,\cite{Hotta2b,Hotta2c}
but due to the complexity of the model including three orbitals,
the mechanism of quasi-Kondo phenomenon due to dynamical
Jahn-Teller phonons has not been completely understood.

In this letter, the Anderson model dynamically coupled with local
Jahn-Teller phonons is numerically analyzed.
Here spin degree of freedom is suppressed for simplicity,
but orbital degree of freedom plays a role of pseudo-spin.
It is shown that total angular moment $J$,
composed of electron pseudo-spin and phonon angular moments,
is screened by conduction electrons,
leading to the singlet ground state of $J=0$.
A characteristic temperature $T_{\rm K}$ defined by a peak
in the specific heat is discussed on the basis of the $s$-$d$
Hamiltonian derived from the Jahn-Teller-Anderson model
in the non-adiabatic region.
To complete the discussion,
we also consider $T_{\rm K}$ in the adiabatic region.


Let us introduce the Anderson model coupled with
local Jahn-Teller phonons.
In order to focus on the effect of coupling between electrons and
local phonons, we consider active orbital degree of freedom,
but for simplicity, spin degree of freedom is suppressed.
Then, the model is expressed as~\cite{unit}
\begin{equation}
  H=\sum_{\mib{k}\tau} \varepsilon_{\mib{k}}
    c_{\mib{k}\tau}^{\dag} c_{\mib{k}\tau}
   +\sum_{\mib{k}\tau} (Vc_{\mib{k}\tau}^{\dag}d_{\tau}+{\rm h.c.})
   +H_{\rm loc},
\end{equation}
where $\varepsilon_{\mib{k}}$ is the dispersion of conduction electron,
$c_{\mib{k}\tau}$ is an annihilation operator of conduction electron
with momentum $\mib{k}$ and orbital $\tau$,
$d_{\tau}$ is an annihilation operator of localized electron
with orbital $\tau$ on an impurity site,
and $V$ is the hybridization between conduction and localized electrons.
We set $V=0.25$ and the energy unit is a half of the conduction
bandwidth, $D$, which is set as 1 eV throughout this paper.

The local electron term $H_{\rm loc}$ is given by
\begin{equation}
  H_{\rm loc}=Un_an_b+\mu\rho+H_{\rm eph},
\end{equation}
where $U$ denotes inter-orbital Coulomb interaction,
$n_{\tau}=d^{\dag}_{\tau}d_{\tau}$, $\rho=n_a+n_b$,
and $\mu$ is a chemical potential.
In this paper, we consider the symmetric case with $\mu=-U/2$.
The electron-phonon coupling term $H_{\rm eph}$ is given by
\begin{equation}
  \begin{split}
    H_{\rm eph} &= g (Q_2 \tau_x + Q_3 \tau_z)
     +(P_2^2+P_3^2)/2 \\
    &+\omega^2(Q_2^2+Q_3^2)/2,
  \end{split}
\end{equation}
where $g$ is the electron-phonon coupling constant,
$Q_2$ and $Q_3$ are normal coordinates for $(x^2-y^2)$-
and $(3z^2-r^2)$-type Jahn-Teller phonons, respectively,
$P_2$ and $P_3$ are corresponding canonical momenta,
$\tau_{x}=d_{a}^{\dag}d_{b}+d_{b}^{\dag}d_{a}$,
$\tau_{z}=d_{a}^{\dag}d_{a}-d_{b}^{\dag}d_{b}$,
and $\omega$ is the frequency of local Jahn-Teller phonons.
Note that the reduced mass is set as unity.
When we introduce phonon operators $a_2$ and $a_3$ through
$Q_2=(a_2+a_2^{\dag})/\sqrt{2\omega}$ and
$Q_3=(a_3+a_3^{\dag})/\sqrt{2\omega}$,
$H_{\rm eph}$ is rewritten as
\begin{equation}
  \begin{split}
  H_{\rm eph} &= \omega\sqrt{\alpha}
  [(a_2+a_2^{\dag}) \tau_x + (a_3+a_3^{\dag}) \tau_z] \\
  &+ \omega(a_2^{\dag}a_2+a_3^{\dag}a_3+1),
  \end{split}
\end{equation}
where $\alpha$ is the non-dimensional electron-phonon coupling constant,
defined as $\alpha=g^2/(2\omega^3)$.


Here we consider the conserved quantity of $H_{\rm eph}$.\cite{Takada}
Using the commutation relation
$[ \tau_{\alpha}, \tau_{\beta} ]=
2{\rm i}\epsilon_{\alpha\beta\gamma}\tau_{\gamma}$
with the Levi-Civita tensor $\epsilon$,
we obtain $[H_{\rm eph}, J]=0$,
where $J$ denotes total angular moment,
defined by $J=L_y+\tau_y/2$ with
$L_y=-{\rm i}(a_3^{\dag}a_2-a_2^{\dag}a_3)$
and
$\tau_{y}=-{\rm i}(d_a^{\dag}d_b-d_b^{\dag}d_a)$.
It is convenient to introduce the electron and phonon bases
to diagonalize $\tau_y$ and $L_y$, respectively.
Namely, we introduce
$b_{\pm}^{\dag}=(a_3^{\dag} \pm {\rm i}a_2^{\dag})/\sqrt{2}$
and
$d_{\sigma}^{\dag}=(d_{a}^{\dag} \pm {\rm i}d_{b}^{\dag})/\sqrt{2}$,
where $\sigma$ denotes a pseudospin and
$\uparrow$ ($\downarrow$) corresponds to $+$($-$) sign.
Then, we obtain
\begin{equation}
  \label{Hloc}
  \begin{split}
  H_{\rm loc} &= Un_{\uparrow}n_{\downarrow} + \mu\rho
   + \omega(b_{+}^{\dag}b_{+}+b_{-}^{\dag}b_{-}+1) \\
  &+ \omega\sqrt{2\alpha}
    [(b_{+} + b_{-}^{\dag}) \sigma_{+}
    +(b_{-} + b_{+}^{\dag}) \sigma_{-}],
  \end{split}
\end{equation}
where $n_{\sigma}=d_{\sigma}^{\dag}d_{\sigma}$,
$\rho=n_{\uparrow}+n_{\downarrow}$,
$\sigma_{+}=d_{\uparrow}^{\dag} d_{\downarrow}$,
and
$\sigma_{-}=d_{\downarrow}^{\dag} d_{\uparrow}$.
The total angular moment $J$ is given by
\begin{eqnarray}
  J=L_z+\sigma_z/2,
\end{eqnarray}
where $L_{z}=b_{+}^{\dag}b_{+}-b_{-}^{\dag}b_{-}$ and
$\sigma_z=d_{\uparrow}^{\dag}d_{\uparrow}-d_{\downarrow}^{\dag}d_{\downarrow}$.
Corresponding to the above transformations, we also introduce
$c_{\mib{k}\sigma}^{\dag}=
(c_{\mib{k}a}^{\dag} \pm {\rm i}c_{\mib{k}b}^{\dag})/\sqrt{2}$
for conduction electrons.
Then, the whole Hamiltonian is still expressed as eq.~(1),
when we replace orbital index $\tau$ with pseudo-spin $\sigma$.


Let us briefly discuss the vibronic state of $H_{\rm loc}$ at half-filling.
The $k$th eigen-state of $H_{\rm loc}$ labeled by $J$ is given by
\begin{equation}
  \label{eqloc}
  |\Phi_{\rm loc}^{(k,J)} \rangle \!=\!
  \sum_{n=0}^{\infty} \Bigl[
   p_{k,n}^{(J)}d^{\dag}_{\uparrow}|J \!-\! \frac{1}{2}; n \rangle
  +q_{k,n}^{(J)}d^{\dag}_{\downarrow} |J \!+\! \frac{1}{2}; n \rangle
  \Bigr],
\end{equation}
where $|L; n \rangle=|L+n, n \rangle$ for $L \ge 0$
and $|n, n+|L| \rangle$ for $L<0$ with
$|n_+,n_- \rangle$=
$(1/\sqrt{n_+! n_-!})(a_+^{\dag})^{n_+}(a_-^{\dag})^{n_-}|0\rangle$
and the vacuum state $|0\rangle$.
The corresponding eigen-energy is given by $E^{(k,J)}_{\rm loc}+\mu$.
Note that $J$ takes half-odd-integer values as
$J=\pm 1/2$, $\pm 3/2$, $\pm 5/2$, $\cdots$
and the coefficients satisfy the relations of
$p_{k,n}^{(-J)}=q_{k,n}^{(J)}$ and
$q_{k,n}^{(-J)}=p_{k,n}^{(J)}$.
The double degeneracy in each eigen-state originates from
the rotational mode of Jahn-Teller phonons.
The ground state is specified by $J=\pm 1/2$ and $k=1$.
It is clearly observed that the phonon state with $L_z=J+1/2$ ($J-1/2$)
is coupled with the electron state with pseudo-spin $S_z=-1/2$ ($+1/2$),
leading to the vibronic state specified by total angular momentum $J$.


In order to investigate electronic and phononic properties of $H$
at low temperatures, we usually discuss corresponding susceptibilities.
The susceptibility of an arbitrary operator $A$ is expressed by
\begin{equation}
  \label{sus}
  \chi_{A}=\frac{1}{Z} \sum_{n,m}
  \frac{e^{-E_n/T}-e^{-E_m/T}}{E_m-E_n}
  | \langle m | A | n \rangle|^2,
\end{equation}
where $E_n$ is the eigen-energy for the $n$th eigenstate
$|n\rangle$ of $H$ and $Z$ is the partition function given by
$Z=\sum_n e^{-E_n/T}$.
For the evaluation of susceptibilities, here we employ
a numerical renormalization group (NRG) method \cite{NRG},
in which momentum space is logarithmically discretized
to include efficiently the conduction electrons near the Fermi energy
and the conduction electron states are characterized by ``shell''
labeled by $N$.
The shell of $N=0$ denotes an impurity site described by
the local Hamiltonian.
The Hamiltonian is transformed into the recursion form as
\begin{eqnarray}
  H_{N+1} = \sqrt{\Lambda} H_N+t_N \sum_{\sigma}
  (c_{N\sigma}^{\dag}c_{N+1\sigma}+c_{N+1\sigma}^{\dag}c_{N\sigma}),
\end{eqnarray}
where $\Lambda$ is a parameter for logarithmic discretization,
$c_{N\sigma}$ denotes the annihilation operator of conduction electron
in the $N$-shell, and $t_N$ indicates ``hopping'' of electron between
$N$- and $(N+1)$-shells, expressed by
\begin{eqnarray}
  t_N=\frac{(1+\Lambda^{-1})(1-\Lambda^{-N-1})}
  {2\sqrt{(1-\Lambda^{-2N-1})(1-\Lambda^{-2N-3})}}.
\end{eqnarray}
The initial term $H_0$ is given by
\begin{equation}
  H_0=\Lambda^{-1/2} [H_{\rm loc}
  +\sum_{\sigma}V(c_{0\sigma}^{\dag}d_{\sigma}+d_{\sigma}^{\dag}c_{0\sigma})].
\end{equation}
We also evaluate entropy $S_{\rm imp}$ and specific heat $C_{\rm imp}$
of localized electron.
In the NRG calculations, a temperature $T$ is defined as
$T=\Lambda^{-(N-1)/2}$.
In this paper, we set $\Lambda$=$2.5$ and we keep $M=5000$
low-energy states for each renormalization step.
The phonon basis for each Jahn-Teller mode is truncated
at a finite number $N_{\rm ph}$, which is set as $N_{\rm ph}=30$.


Before proceeding to the NRG results,
it is useful to examine the vibronic state of $H_0$.
The $k$th eigen-state of $H_0$ is found to be expressed as~\cite{note}
\begin{equation}
 \begin{split}
  |\Phi_0^{(k,J)} \rangle & \!=\! \sum_{n=0}^{\infty}
    \Bigl[
     \frac{a_{k,n}^{(J)}}{\sqrt{2}}
    (c^{\dag}_{0\uparrow}c^{\dag}_{0\downarrow}
    +d^{\dag}_{\uparrow}d^{\dag}_{\downarrow})|J; n\rangle \\
   &+\frac{b_{k,n}^{(J)}}{\sqrt{2}}
   (c^{\dag}_{0\uparrow}d^{\dag}_{\downarrow}
   -d^{\dag}_{\uparrow}c^{\dag}_{0\downarrow})|J; n\rangle \\
   &+\frac{c_{k,n}^{(J)}}{\sqrt{2}}
   (c^{\dag}_{0\uparrow}d^{\dag}_{\uparrow}|J-1; n\rangle
   -c^{\dag}_{0\downarrow}d^{\dag}_{\downarrow}|J+1; n\rangle)
   \Bigr],
 \end{split}
\end{equation}
where $J$ takes an integer.
Note that the ground state is characterized by $J=0$ and $k=1$.

\begin{figure}[t]
\begin{center}
\includegraphics[width=8.5truecm]{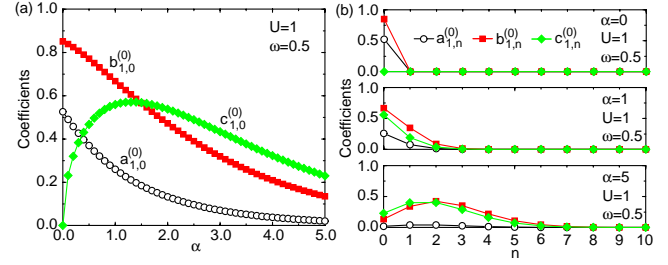}
\caption{(Color online)
(a) Coefficients in $|\Phi_0^{(1,0)}\rangle$ vs. $\alpha$
for $U=1$ and $\omega=0.5$.
(b) Coefficients in $|\Phi_0^{(1,0)}\rangle$ vs. $n$
for $\alpha=0$, $1$, and $5$.
}
\end{center}
\end{figure}

In Fig.~1(a), we plot the coefficients $a_{1,0}^{(0)}$,
$b_{1,0}^{(0)}$, and $c_{1,0}^{(0)}$ vs. $\alpha$
in the ground state for $U=1$ and $\omega=0.5$.
At $\alpha=0$, the system is described by the Anderson model
and we find the singlet ground state in which total spin is zero.
When we increase $\alpha$, the component of $c_{1,0}^{(0)}$
is rapidly increased.
Namely, the phonon state with $L_z=+1$ ($-1$) is
coupled to the electron state with total pseudo-spin
$S_z^{\rm tot}=-1$ ($+1$),
leading to the singlet state with $J=0$.
When we further increase $\alpha$, multi-phonon components
with $n \ge 1$ become significant, as shown in Fig.~1(b).
The broad peaks in the coefficients are
shifted to larger $n$ with increasing $\alpha$.
In short, the singlet ground state with $J=0$
includes the significant component of phonon state
with the moment $L_z = \pm 1$, which is cancelled by the pseudo-spin
moment of the electron state with $S_z^{\rm tot}=\mp 1$.
It is emphasized that phonon degrees of freedom essentially
contribute to the formation of the singlet ground state,
which is an important point of the Kondo effect in
an electron system with dynamical Jahn-Teller impurity.


Let us now discuss the NRG results of the Jahn-Teller-Anderson model.
In Fig.~2(a), we show entropy and specific heat
for $\alpha=0.3$, $\omega=0.5$, and $U=1$.
An entropy of $\log 2$ originates from the double degeneracy
in the local vibronic state with $J=\pm 1/2$,
corresponding to clockwise and anti-clockwise rotational modes
of dynamical Jahn-Teller phonons.\cite{Hotta1}
Here we define $T^*$ as a temperature at which the entropy becomes
$\log 2$. In Fig.~2(a), we obtain $T^* \sim 0.03$.
When we decrease temperature,
the entropy of $\log 2$ is eventually released and
we observe a peak in the specific heat,
which defines a characteristic temperature $T_{\rm K}$.
In Fig.~2(a), we obtain $T_{\rm K} \sim 3\times 10^{-6}$.
The Kondo phenomenon due to dynamical Jahn-Teller phonons
is clearly observed for the case of $T_{\rm K} \ll T^*$,
which is satisfied in the non-adiabatic region.

In Fig.~2(b), the results of several kinds of susceptibilities
are shown.
We find that $\chi_{\sigma_x}=\chi_{\sigma_y}>\chi_{\sigma_z}$,
where $\sigma_x=\sigma_{+}+\sigma_{-}$ and
$\sigma_y=-{\rm i}(\sigma_{+}-\sigma_{-})$,
due to the effect of Jahn-Teller phonons coupled with
quadrupole operators $\tau_z$ and $\tau_x$,
which are $\sigma_x$ and $\sigma_y$, respectively,
in the pseudo-spin notation.
We remark that charge susceptibility immediately vanishes
at high temperatures,
indicating that charge degree of freedom is irrelevant to
the present Kondo-like phenomenon.
On the other hand, pseudo-spin susceptibilities and $\chi_J$ are
suppressed around at $T=T_{\rm K}$.
Due to the discussion on the vibronic state of $H_0$,
we deduce that the total moment $J$, composed of pseudo spin and
phonon angular moments, is screened to form the singlet
ground state of $J=0$.
In order to confirm that the present behavior is really
the Kondo effect concerning $J$,
we check that $\chi_J$ is scaled by a single parameter $T_{\rm K}$.
In Fig.~2(c), we plot $T\chi_J$ as a function of $T/T_{\rm K}$
for $\alpha=0.1 \sim 0.7$.
We actually find that $T\chi_J$ is expressed by a single function
of $T/T_{\rm K}$ within the numerical precision,
except for the high-temperature region in which the renormalization
is not enough.
Note that the pseudo-spin susceptibility does not satisfy
such a scaling relation.

In order to understand the behavior of $T_{\rm K}$,
it is useful to derive an effective model from $H$ by using
the perturbation expansion in terms of $V$.
Note that the condition for the expansion is given by
$\omega \gg 2\Delta$,
where $\Delta=\pi\rho_0V^2$ is the width of the virtual bound state
and $\rho_0$ is the density of state of the conduction electron
at the Fermi level.
Namely, the following effective model is valid
in the non-adiabatic region with large phonon energy.
The adiabatic region will be discussed later.

\begin{figure}[t]
\begin{center}
\includegraphics[width=8.5truecm]{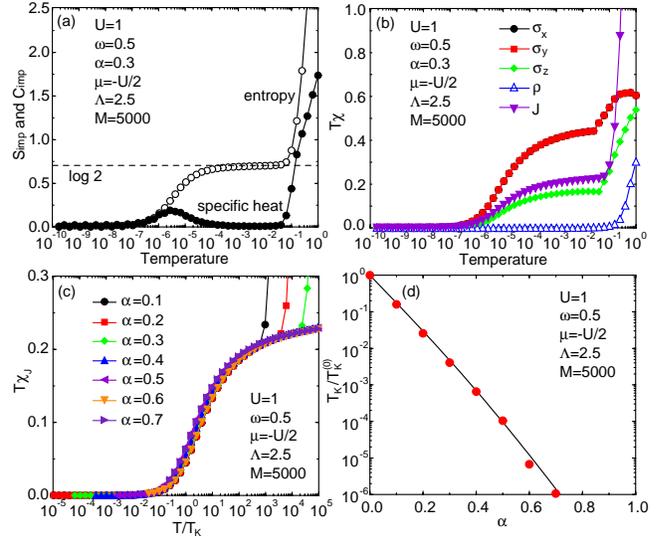}
\caption{(Color online)
(a) Entropy and specific heat for $\alpha=0.3$, $\omega=0.5$, and $U=1$.
(b) Susceptibilities for pseudo spin, charge, and total
angular moment for the same parameters as in (a).
(c) $T\chi_J$ as a function of the scaled temperature $T/T_{\rm K}$
for $\alpha=0.1 \sim 0.7$.
(d) $T_{\rm K}/T_{\rm K}^{(0)}$ vs. $\alpha$ for $U=1$ and $\omega=0.5$.
Solid circles denote the numerical results, while
a solid curve indicates the analytic result of the $s$-$d$ model.
}
\end{center}
\end{figure}

Here we distinguish the degenerate ground state of $H_{\rm loc}$ as
$|{\rm R}\rangle=|\Phi_{\rm loc}^{(1,1/2)} \rangle$
and
$|{\rm L}\rangle=|\Phi_{\rm loc}^{(1,-1/2)} \rangle$,
where ``R'' and ``L'' denote the rotational directions
and the ground-state energy is given by $E^{(1,1/2)}_{\rm loc}+\mu$.
After some algebraic calculations, we obtain the effective
$s$-$d$ model as
\begin{equation}
 \begin{split}
  H_{\rm eff} &= \sum_{\mib{k}\sigma} \varepsilon_{\mib{k}}
  c_{\mib{k}\sigma}^{\dag} c_{\mib{k}\sigma}
  +\sum_{{\mib k},{\mib k'}}
  [J_z(c_{\mib{k}\uparrow}^{\dag} c_{\mib{k'}\uparrow}
  -c_{\mib{k}\downarrow}^{\dag} c_{\mib{k'}\downarrow})S_z \\
  &+J_{\bot}(c_{\mib{k}\downarrow}^{\dag} c_{\mib{k'}\uparrow}S_+
  +c_{\mib{k}\uparrow}^{\dag} c_{\mib{k'}\downarrow}S_-)],
 \end{split}
\end{equation}
where
$S_z=(|{\rm R} \rangle \langle {\rm R}|-|{\rm L}\rangle \langle {\rm L}|)/2$,
$S_{+}=|{\rm R} \rangle \langle {\rm L} |$,
$S_{-}=|{\rm L} \rangle \langle {\rm R} |$,
and the exchange interactions are given by
\begin{equation}
 J_{\bot} = \sum_{n=0}^{\infty}
   \frac{2V^2 [p_{1,n}^{(1/2)}]^2}
   {U/2-E^{(1,1/2)}_{\rm loc}+\omega (2n+1)},
\end{equation}
and
\begin{equation}
 J_z=J_{\bot}-\sum_{n=0}^{\infty}
   \frac{2V^2 [q_{1,n}^{(1/2)}]^2}
   {U/2-E^{(1,1/2)}_{\rm loc}+\omega (2n+2)}.
\end{equation}
Note here that $J_{\bot}>J_z$ for $\alpha>0$.
Namely, the significant effect of non-adiabatic phonons appears in
the enhancement of the transverse component of the exchange interaction.
This is consistent with $\chi_{\sigma_x}=\chi_{\sigma_y}>\chi_{\sigma_z}$
for $T>T_{\rm K}$ in Fig.~2(b), since Jahn-Teller phonons are coupled
with $\sigma_{+}$ and $\sigma_{-}$.
In other words, the directions of the rotational Jahn-Teller mode
are easily converted for high-energy phonons.

For the $s$-$d$ model with anisotropic exchange interaction,
Shiba has obtained the explicit expression for
the binding energy ${\tilde E}$.\cite{Shiba}
When we define the Kondo temperature $T_{\rm K}$ as
$T_{\rm K}=-{\tilde E}$, we obtain $T_{\rm K}$ as~\cite{Shiba}
\begin{equation}
 \label{Tk}
 T_{\rm K}=D {\rm exp} \Biggl[
 \frac{-1}{2\rho_0\sqrt{J_{\bot}^2-J_z^2}}
 \tan^{-1}\Bigl(\frac{\sqrt{J_{\bot}^2-J_z^2}}{J_z} \Bigr) \Biggr].
\end{equation}
In Fig.~2(d), we depict $T_{\rm K}/T_{\rm K}^{(0)}$ vs. $\alpha$,
where $T_{\rm K}^{(0)}$ is the Kondo temperature of
the Anderson model with $\alpha=0$.
Numerical results are shown by solid symbols.
Note that $T_{\rm K}$ as well as $T_{\rm K}^{(0)}$
in the numerical results is defined as a temperature
which shows the peak in the specific heat.
Analytic result is depicted by a solid curve,
which indicates eq.~(\ref{Tk}) divided by
$T_{\rm K}^{(0)}=De^{-1/(2J_0\rho_0)}$ with $J_0=4V^2/U$.
We find that the numerical results agree well with
the analytic curve for the $s$-$d$ model with
the anisotropic exchange interaction.

\begin{figure}[t]
\begin{center}
\includegraphics[width=8.5truecm]{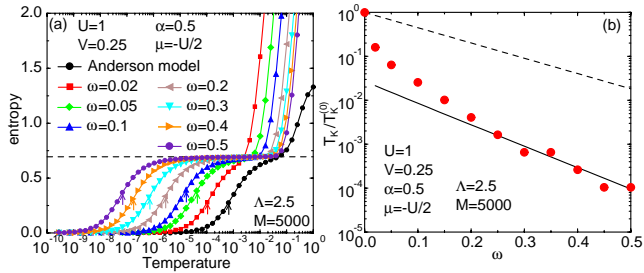}
\caption{(Color online)
(a) Entropies for $\omega=0.02 \sim 0.5$ with $U=1$ and $\alpha=0.5$.
Arrows denote the Kondo temperatures defined by the peaks in
the specific heat.
(b) $T_{\rm K}/T_{\rm K}^{(0)}$ vs. $\omega$ for $U=1$ and $\alpha=0.5$.
Solid symbol denotes the numerical results.
Solid and broken curves indicate the analytic results
for non-adiabatic and adiabatic cases, respectively.
}
\end{center}
\end{figure}


Thus far, we have concentrated on the non-adiabatic region,
since we are interested in the Kondo effect due to
{\it dynamical} Jahn-Teller phonons,
but in order to complete the discussion,
let us also discuss the adiabatic region.
In Fig.~3(a), we show entropies for several values
of $\omega$ with $U=1$ and $\alpha=0.5$.
When $\omega$ is decreased, $T^*$ is decreased and
it gradually approaches $T_{\rm K}^{(0)}$.
In Fig.~3(b), $T_{\rm K}/T_{\rm K}^{(0)}$ is plotted
as a function of $\omega$ for $U=1$ and $\alpha=0.5$.
For $\omega>0.2$, the numerical results are well reproduced
by the analytic expression of the $s$-$d$ model,
since $T^*$ and $T_{\rm K}^{(0)}$ are clearly separated.
However, for smaller $\omega$, the numerical results
are significantly deviated from the solid curve of eq.~(\ref{Tk}).
In the adiabatic region, as shown in Fig.~3(a),
$T^*$ becomes comparable with $T_{\rm K}^{(0)}$ and
the effective $s$-$d$ model for
non-adiabatic phonons loses its physical meaning.

For adiabatic Jahn-Teller phonons,
we intuitively understand the decrease of $T_{\rm K}$
with increasing $\alpha$ due to the effective enhancement of
$U$ by the static Jahn-Teller energy $E_{\rm JT}$.
\cite{Nagaosa,Hotta3,Kim}
In fact, the Kondo temperature in the adiabatic region is
given by $T^{\rm ad}_{\rm K}=De^{-1/(2J_{\rm eff}\rho_0)}$,
where $J_{\rm eff}=4V^2/(U+4E_{\rm JT})$
with $E_{\rm JT}=\alpha \omega$.
In Fig.~3(b), we show $T^{\rm ad}_{\rm K}/T_{\rm K}^{(0)}$
by a broken curve.
It is considered that the numerical results should be gradually
changed from the solid to the broken curve with decreasing $\omega$.
In order to observe such a gradual change more clearly,
it is necessary to carry out the calculations
for $\omega$ less than $0.01$.
However, in the present value of $M=5000$,
it is quite difficult to perform the NRG calculations for $\omega<0.01$
with sufficient precision, since low-energy phonon excited
states cannot be fully included.
For the purpose to include the adiabatic phonons effectively,
it is necessary to improve significantly the NRG calculations,
not by simply increasing the value of $M$.
For instance, it may be important to include
phonon excitations as the polaron effect.
We postpone such effort as one of future problems.


Throughout this paper, we have ignored spin degree of freedom.
In spite of this simplification, we have found the interesting
phenomenon that Jahn-Teller phonons contribute to the singlet formation.
Note that this Kondo effect is related to orbital degree of freedom.
In actual materials, of course, there should exist active spin degree
of freedom.
In this sense, it is necessary to analyze the two-orbital Anderson model
including Jahn-Teller phonons which are coupled with
orbital (quadrupole) degree of freedom.
In particular, it is an interesting problem to clarify how
the Kondo effect due to dynamical Jahn-Teller phonons
occurs in the spin-orbital system.
It is another future problem.

Since in general, a spin-orbit interaction is strong
for $f$ electrons, readers may think that
it is meaningless to use the model with spin and/or orbital
degrees of freedom for $f$-electron systems.
However, by exploiting the $j$-$j$ coupling scheme,
it is possible to construct microscopic models
for $f$-electron systems,\cite{Hottajj1,Hottajj2,Hottajj3}
in which spin degree of freedom is introduced
to distinguish the degenerate states in the Kramers doublet,
while orbital degree of freedom distinguishes
the different kinds of Kramers doublets.
Thus, it is meaningful to consider the spin-orbital model
for $f$-electron systems.

Note, however, that relevant degree of freedom of $f$ electron
is generally described by {\it multipole}.
In the total angular momentum $J$,
$\tau_y$ is 2u octupole with xyz symmetry and
$L_y$ denotes Jahn-Teller phonon angular moment,
indicating that $J$ does $not$ include the magnetic moment.
Thus, in principle, the Kondo phenomenon due to
dynamical Jahn-Teller phonons is distinguished
from the standard Kondo effect concerning the magnetic moment.
However, the dominant multipole degree of freedom is changed
according to the crystalline electric field (CEF) ground state.
Furthermore, some multipoles are mixed
depending on the crystal structure, CEF parameters,
and $f$-electron number.
The Kondo phenomenon due to dynamical Jahn-Teller phonons
in actual $f$-electron materials
such as Sm-based filled skutterudites
will be discussed in detail elsewhere.\cite{Hotta2c}


In summary, we have studied the Kondo phenomenon
in the Anderson model coupled with dynamical Jahn-Teller phonons.
It has been found that total angular momentum $J$ composed
of the electron pseudo-spin and phonon angular moments
is screened by conduction electrons to form
the singlet ground state of $J=0$.
We have shown that the characteristic temperature is well
explained by the $s$-$d$ model with anisotropic exchange
interaction derived from the Jahn-Teller-Anderson model
in the non-adiabatic region.


The author thanks K. Ueda for discussions.
He is supported by the Japan Society for the Promotion
of Science and by the Ministry of Education, Culture, Sports, Science,
and Technology of Japan.
The computation in this work has been done using the facilities
of the Supercomputer Center of Institute for Solid State Physics,
University of Tokyo.



\begin{thebibliography}{99}
\bibitem{unit}
Throughout this paper, we use such units as $\hbar=k_{\rm B}=1$.
\bibitem{note}
Since we consider the symmetric case with $\mu=-U/2$,
we use the same coefficient for double-occupied states,
$c^{\dag}_{0\uparrow}c^{\dag}_{0\downarrow}|J; n\rangle$
and $d^{\dag}_{\uparrow}d^{\dag}_{\downarrow}|J; n\rangle$.
In general, these two terms have different coefficients.
\end{thebibliography}
\end{document}